\documentclass[preprint,showpacs,preprintnumbers,amsmath,amssymb]{revtex4}

\usepackage{graphicx}
\usepackage{dcolumn}
\usepackage{bm}

\begin{document}


\title{Distortion of Schwarzschild-anti-de Sitter black holes  \\ to black strings} 

\author{Akira Tomimatsu}
\email{atomi@gravity.phys.nagoya-u.ac.jp}
\affiliation{Department of Physics, Nagoya University, Nagoya 464-8602, Japan} 


\date{\today}

\begin{abstract}
Motivated by the existence of black holes with various topologies in four-dimensional spacetimes with a negative cosmological constant, we study axisymmetric static solutions describing any large distortions of Schwarzschild-anti-de Sitter black holes parametrized by the mass $m$. Under the approximation such that $m$ is much larger than the anti-de Sitter radius, it is found that a cylindrically symmetric black string is obtained as a special limit of distorted spherical black holes. Such a prolonged distortion of the event horizon connecting a Schwarzschild-anti-de Sitter black hole to a black string is allowed without violating both the usual black hole thermodynamics and the hoop conjecture for the horizon circumference.
\end{abstract}

\pacs{04.20.-q, 04.70.-s}
\maketitle

\section{\label{sec:level1}INTRODUCTION} 

Stationary black holes in spacetimes with a negative cosmological constant $\Lambda$ (which we call hereafter anti-de Sitter black holes for abbreviation) have been a subject of current interest, in particular, motivated by the conjecture of the anti-de Sitter/Conformal field theory-correspondence \cite{Mal98}, according to which the emergence of black hole thermodynamics is interpreted in terms of thermal states of the dual conformal field theory \cite{Witt1,Witt2}. It is also interesting that the vacuum Einstein equations with the cosmological term admit stationary exact solutions which have the event horizon with various topologies even in four dimensions \cite{Lemos1,Lemos2,Huang,Lemos3,Cai,Ami,Brill,Vanzo97,Klemn98}. Namely, the spatial section of the event horizon can be Einstein's manifold with positive, zero or negative scalar curvature corresponding to $k=+1, 0, -1$. The thermodynamic properties of such anti-de Sitter (AdS) black holes have been investigated in \cite{HP} for $k=+1$ and in \cite{Brill,Vanzo97} for $k=0,-1$. Though the higher-dimensional generalization has been extensively studied (see, for example, \cite{Mann,Bir}), in this paper we focus our attention on the four-dimensional black holes.

For $k=1$ the event horizon has a spherical topology ($S^{2}$), and the Schwarzschild-anti-de Sitter (SAdS) solution is well-known as a typical example of spherically symmetric anti-de Sitter black holes. On the other hand, the horizon topology corresponding to zero scalar curvature (i.e., $k=0$) may be planar ($R^{2}$), cylindrical ($R^{1}\times S^{1}$) or toruslike ($S^{1}\times S^{1}$) according to the compactification scheme for the two-dimensional spatial section. Hyperbolic (sometimes called topological) black holes represented by the $k=-1$ solutions may have a negative mass, and the (in)stability becomes a subtle problem \cite{Gib,Neu}.

It is remarkable that  conformal (compactified) spatial infinity of a stationary AdS black hole spacetime has the same topology as the event horizon. If the horizon topology is fixed, a black hole solution satisfying the stationary vacuum Einstein equations with $\Lambda<0$ may be unique \cite{And,Kod}. However, this uniqueness theorem holds only if the boundary metric at conformal spatial infinity is required to be the Einstein metric with a constant scalar curvature. In fact, axisymmetric static perturbations of the SAdS solution have been explicitly presented in \cite{Yos}. The key result of the perturbative analysis is that a static small distortion from spherical symmetry (which is regular at the event horizon) does not vanish even at spatial infinity. Namely, we obtain a non-uniform spherical 2-surface at spatial infinity, where an arbitrary function $f(\theta)$ dependent on the zenithal angle $\theta$ appears. This should be compared to vacuum black hole (namely, Schwarzschild and Kerr) solutions with $\Lambda=0$, for which any static distortion regular at the horizon must diverge at spatial infinity (see \cite{Ger, Tom} for the exact solutions of distorted black holes). The role of a negative cosmological constant is to give a finite distortion to the boundary metric on $S^{2}$ at spatial infinity. Interestingly, we note the one-to-one correspondence between the boundary data represented by the function $f(\theta)$ and the bulk spacetime geometry from the horizon to spatial infinity.

In this paper we consider a large distortion of a SAdS black hole as an extension of the analysis in \cite{Yos} to a non-perturbative case. Our purpose is to construct a family of distorted anti-de Sitter black hole solutions connecting the $k=1$ (SAdS) black hole solution to the $k=0$ black string solution with a cylindrical horizon topology. This will be useful for studying the important area of AdS black hole physics concerning the quasi-static or thermodynamic evolution which may be accompanied with a large distortion from spherical symmetry.

For mathematical convenience our investigation is limited to the large mass domain such that $m \gg r_{A}$, where $m$ is the SAdS mass parameter and $r_{A}\equiv \sqrt{-3/\Lambda}$ is the anti-de Sitter radius. (Hereafter we use units such that $c=G=\hbar=k_B=1$.) Because in the large mass domain the inequality $r_{A} \ll r_{H} \ll m$ holds for the SAdS horizon radius $r_{H}$, the effect of $\Lambda$ becomes significant even near the horizon, and the horizon 2-surface is distorted in the nearly same manner as the 2-surface at spatial infinity, as was shown in \cite{Yos}. This allows us to obtain easily a non-perturbative distortion of the metric in Sec. II. Then, in Sec. III, we estimate the mass of distorted AdS black holes, using the definition proposed by Ashtekar and Magnon \cite{AM} applicable to spacetimes with a distorted boundary metric at spatial infinity. We find the condition for distortion keeping the Astekar-Magnon mass equal to the SAdS mass parameter $m$. It is also possible to obtain the Euclidean gravitational action, using the boundary counterterm technique \cite{Emp,Mann2}. The leading-order calculation in the large mass domain clearly shows that the thermodynamic mass energy is equal to $m$. The entropy and the temperature are given as functions of $m$ in the same way as the SAdS case. We can conclude that any effect due to the horizon distortion becomes thermally insignificant in the large mass domain. In Sec. IV, we present an explicit example representing a black hole distortion connecting to a uniform black string. Such a distortion is shown to satisfy the so-called Penrose inequality for the mass and the horizon area, of which the validity has been discussed in asymptotically flat spacetimes (see, for example, \cite{Bray}). We also mention, in terms of the hoop conjecture proposed by Thorne \cite{Thorne}, the possible existence of an extremal state of distorted (non-rotating) AdS black holes.

\section{\label{Dis}STATIC AXISYMMETRIC DISTORTION}
Without loss of generality the static axisymmetric metric for describing a black hole with a spherical horizon topology is given by
\begin{equation}
ds^2=-e^{2\nu}dt^2+e^{2\mu}dr^2+r^2e^{2\psi}(d\theta^2+\sin^2\theta d\phi^2) ~ ,
\label{metric}
\end{equation}
where the functions $\nu$, $\mu$ and $\psi$ depend only on $r$ and $\theta$. For the SAdS metric we obtain
\begin{equation}
e^{2\nu_0}=e^{-2\mu_0}=1-\frac{2m}{r}+(\frac{r}{r_{A}})^2 ~ , ~~ e^{2\psi_0}=1 ~ .
\label{sads}
\end{equation}
If we consider a static axisymmetric distortion of the SAdS metric according to a usual perturbative scheme, we have up to the first order
\begin{equation} 
\nu=\nu_0+\epsilon\nu_1~,~~\mu=\mu_0+\epsilon\mu_1~,~~\psi=\epsilon\psi_1~,
\end{equation}
with a small parameter $\epsilon$. Using the vacuum Einstein equations $R_{ab}=\Lambda g_{ab}$ with a negative cosmological constant $\Lambda$, the first-order perturbations $\nu_1(r,\theta)$, $\mu_1(r,\theta)$ and  $\psi_1(r,\theta)$ have been studied in \cite{Yos}. For each multiple component given by Legendre's polynomial $P_l(\cos\theta)$ with $\l\geq 2$ we obtain
\begin{equation}
\nu_1=-\mu_1=-H^{(1)}(r)P_l(\cos\theta)~,~~\psi_1=K^{(1)}(r)P_l(\cos\theta)~.
\end{equation}
In this paper we focus our interest on the large mass domian such that $m\gg r_{A}$, in which the radial functions are given by
\begin{equation}
H^{(1)}=\frac{3x}{x^2+x+1}-\frac{1}{x}+O(\delta)~,~~K^{(1)}=\frac{6}{\delta(l^2+l-2)}+O(1)~,
\label{radial}
\end{equation}
where $\delta\equiv r_H/2m\ll 1$,  $x\equiv r/r_H$, and the horizon radius $r_H$ is approximately written by $r_H\simeq(2mr_A^2)^{1/3}$ (see \cite{Yos} for the derivation). The key point of the result (\ref{radial}) is that the functional form of $H^{(1)}$ does not depend on $l$ in the leading-order calculation with respect to the new small parameter $\delta$. This allows us to write the first-order perturbation $\nu _1$ as follows,
\begin{equation}
\nu_1=-\mu_1=(\frac{3x}{x^2+x+1}-\frac{1}{x})h(\theta) ~,
\label{nu1}
\end{equation}
where $h(\theta)$ is an arbitrary function of $\theta$. Note also that the radial function $K^{(1)}$ does not depend on $r$, though the constant value depends on $l$. The first-order perturbation $\psi_1$ can be written by
\begin{equation}
\psi_1=\frac{1}{\delta}f(\theta) ~,
\label{psi1}
\end{equation}
where the function $f(\theta)$ must satisfy the relation
\begin{equation}
6h=\frac{d^2f}{d\theta^2}+\cot\theta\frac{df}{d\theta}+2f~.
\label{linear}
\end{equation}
This linear analysis clearly shows that one arbitrary function (i.e., $f$ or $h$) of $\theta$ appears as a hair of distorted AdS black holes. In particular, for SAdS black holes with large mass $m$, any axisymmetric distortion of a spherical 2-surface ($S^2$) is represented by the function $f(\theta)$ independent of the radial coordinate $r$. Namely, we obtain the same distortion of $S^2$ in the whole range from the horizon $r=r_H$ to spatial infinity $r\rightarrow\infty$.

Now we consider a non-perturbative extension of the results obtained by the linear analysis. It is clear from Eqs. (\ref{nu1}) and (\ref{psi1}) that if the distortion parameter $\epsilon$ is chosen to be $\epsilon=\delta\ll 1$, the linear approximation for $\epsilon\psi_1$ breaks down, while the perturbation $\epsilon\nu_1=-\epsilon\mu_1$ remains small. Hence, the expansion of the metric (\ref{metric}) with respect to the small parameter $\delta$ will allow us to give a non-perturbative distortion only to the metric function $\psi$ defined on $S^2$. 

Note that the SAdS metric (\ref{sads}) can be written by
\begin{equation}
e^{2\nu_0}=e^{-2\mu_0}=\frac{1}{\delta}(x^2-\frac{1}{x})+1
\label{sads2}
\end{equation}
in the range $x\geq 1$. Then the expansion of the metric function $\nu$ with respect to $\delta$ (i.e., $\nu=\nu_0+\delta\nu_1$) should be rewritten into the form
\begin{equation}
e^{2\nu}\simeq e^{-2\mu}\simeq \frac{1}{\delta}(x^2-\frac{1}{x})\{1+2\delta\nu_1+\delta(x^2-\frac{1}{x})^{-1}\}~.
\label{h}
\end{equation}
The approximation such that $e^{2\delta\nu_1}\simeq 1+2\delta\nu_1$ is still possible in the same way as the linear analysis. However, the leading term for the metric function $\psi$ should be
\begin{equation}
e^{2\psi}\simeq e^{2f(\theta)}~,
\label{f}
\end{equation}
which does not allow the approximation such that $e^{2\delta\psi_1}\simeq 1+2\delta\psi_1$. Nevertheless it is straightforward to see that the vacuum Einstein equations with a negative cosmological constant are satisfied up to the next-to leading-order calculation in the small $\delta$ domain, only if the nonlinear relation
\begin{equation}
6h-1=e^{-2f}(\frac{d^2f}{d\theta^2}+\cot\theta\frac{df}{d\theta}-1)~,
\label{basic}
\end{equation}
is required instead of Eq. (\ref{linear}). Of course, for a small distortion corresponding to the approximation $|f|\ll 1$ Eq. (\ref{basic}) reduces to Eq. (\ref{linear}). 

We can treat any non-perturbative distortion of $S^2$ by the use of the arbitrary metric function $f(\theta)$, which also determines the small correction $\delta h(\theta)$ to the metric function $\nu=-\mu$ through Eq. (\ref{basic}). If such a distortion is required to preserve the area of $S^2$, the function $f$ should satisfy the additional condition
\begin{equation}
\frac{1}{2}\int_0^{\pi} e^{2f(\theta)} \sin\theta d\theta=1 ~,
\label{condition}
\end{equation}
from which Eq. (\ref{basic}) leads to the result that the mean value $\bar{h}$ vanishes if the distortion perturbation $h(\theta)$ is averaged over $S^2$, namely, 
\begin{equation}
\bar{h} \equiv \frac{1}{2}\int_0^{\pi} e^{2f(\theta)} h(\theta)\sin\theta d\theta=0 ~.
\label{condition2}
\end{equation}
Let us remark that the area-preserving condition (\ref{condition}) is allowed without loss of generality. For some choice of $f(\theta)$ the integral in Eq. (\ref{condition}) may become equal to $e^{2\bar{f}}$ with a nonzero value $\bar{f}$. Then, we can consider the new distortion functions $f'(\theta)$ and $h'(\theta)$, using the transformations $f \rightarrow f'=f-\bar{f}$ and $h \rightarrow h'=e^{2\bar{f}}h+(1-e^{2\bar{f}})/6$. It is easy to check that these functions $f'$ and $h'$ can satisfy Eqs. (\ref{condition}) and (\ref{condition2}). Further, as was previously mentioned, for a small distortion of $S^2$ (i.e., for $|f|\ll 1$) the distortion functions $f(\theta)$ and $h(\theta)$ may be given by Legendre's polynomial $P_l(\cos\theta)$ with $l\geq 2$. We note that if the distortion is treated as a linear perturbation, Eqs. (\ref{condition}) and (\ref{condition2}) should be regarded as the necessary conditions for $f$ and $h$. Hence, in the following section we will discuss mass and thermodynamic properties of distorted AdS black holes under the condition (\ref{condition}).

\section{THE ASHTEKAR-MAGNON MASS AND THE ENTROPY}
The distortion given by $f(\theta)$ may induce a black hole mass different from the SAdS mass parameter $m$. Though the mass is a basic quantity characterizing black hole states, the definition in spacetimes with a negative cosmological constant remains ambiguous as a problem to be investigated from various viewpoints. If the distortion from sherical symmetry vanishes at spatial infinity, one may use the so-called Abbott-Desser mass $M_{AD}$ \cite{AD}. Unfortunately, as was shown in \cite{Yos}, the calculation based on the Abbott-Deser method which depends on the choice of the background metric is not applicable to distorted AdS black holes. Hence, we adopt here the evaluation method proposed by Ashtekar and Magnon \cite{AM}, for which any background subtraction is unnecessary. The Ashtelar-Magnon mass denoted by $M_{AM}$ is a conserved quantity defined in the conformally transformed spacetime with the metric $\bar{g}_{ab}=\Omega^2g_{ab}$, where $g_{ab}$ is the physical metric (\ref{metric}) representing a AdS black hole. If the conformal factor $\Omega$ is chosen to be $\Omega=1/r$, it is given by
\begin{equation}
M_{AM}=-\frac{r_A}{4}\lim_{r\rightarrow\infty}\int_0^{\pi}\xi^tr^2e^{\nu}
[e^{-2\mu}\{\nu,_{rr}+\nu,_r(\nu,_r-\mu,_r)\}+\frac{e^{-2\psi}}{r^2}\nu,_{\theta}\mu,_{\theta}-\frac{1}{r_A^2}]e^{2\psi}\sin\theta d\theta~,
\label{AMmass}
\end{equation}
where $\xi^t$ is a time component of the timelike Killing vector $\xi^a$ on the conformal boundary at $r\rightarrow\infty$. 

For distorted black holes obtained under the approximation $\delta\ll1$ in the previous section, we find that $\nu_1=-\mu_1\rightarrow0$ in the limit $r\rightarrow\infty$, and the conformal boundary metric is given by
\begin{equation}
d\bar{s}^2=-\frac{1}{r_A^2}dt^2+e^{2f(\theta)}(d\theta^2+\sin^2\theta d\phi^2)~.
\end{equation}
Though the normalization of the Killing vector $\xi^a$ on the boundary metric distorted by $f$ has been discussed in \cite{Yos}, here we choose $\xi^t=1$ in the same way as the SAdS spacetime, considering the condition (\ref{condition}) for $f$. To calculate Eq. (\ref{AMmass}) to the leading order in the small $\delta$ domain, it is sufficient to use the approximated SAdS form
\begin{equation}
e^{2\nu}\simeq e^{-2\mu}\simeq \frac{1}{\delta}(x^2-\frac{1}{x})
\label{sadsapp}
\end{equation}
for $\nu$ and $\mu$ in the integrand, while we have $\psi=f\neq 0$ for the metric function on $S^2$. Then, the final result is simply
\begin{equation}
M_{AM}=m ~.
\end{equation}
In spite of a large distortion of black hole geometry represented by $f$ a change of the black hole mass can remain very small, namely, $(M_{AM}-m)/m=O(\delta)$. 

Let us also check the effect of black hole distortion at the thermodynamic level. For this purpose we calculate the Euclidean gravitational action $I$, using the counterterm prescription applicable to spacetimes with a negative cosmological constant \cite{Emp,Mann2}. For the leading-order calculation of thermodynamic quantities in the small $\delta$ domain, the gravitational action $I$ may be evaluated by giving the Euclidean version of the approximate metric (\ref{sadsapp}) such as
\begin{equation}
ds^2=(\frac{r^2}{r_A^2}-\frac{2m}{r})d\tau^2+(\frac{r^2}{r_A^2}-\frac{2m}{r})^{-1}dr^2
+r^2e^{2f(\theta)}(d\theta^2+\sin^2\theta d\phi^2) ~ ,
\label{metric2}
\end{equation}
where $\tau=it$ is the Euclidean time with the period $\beta$ given by
\begin{equation}
\beta=\frac{2\pi}{3}(\frac{4r_A^4}{m})^{1/3}~.
\end{equation}

The Euclidean gravitational action $I$ contains three contributions denoted by
\begin{equation}
I=I_{bulk}+I_{surf}+I_{ct} ~.
\label{action}
\end{equation}
The first two terms $I_{bulk}$ and $I_{surf}$ in Eq. (\ref{action}) are the familiar classical action corresponding to the volume integral in the range $r_{H}\leq r\leq r_{0}$ and the boundary integral at $r=r_{0}$, respectively, and we have
\begin{equation}
I_{bulk}=\frac{3}{8\pi r_A^2}\int d^4x\sqrt{g}=\frac{\beta}{2r_A^2}(r_0^3-r_H^3)~,
\end{equation}
and
\begin{equation}
I_{surf}=-\frac{1}{8\pi}\int d^3x\sqrt{h}K=\frac{3\beta}{2r_A^2}(-r_0^3+\frac{r_H^3}{2})~,
\end{equation}
where $K$ is the trace of the extrinsic curvature of the boundary with the metric $h_{ab}$, giving the determinant 
\begin{equation}
\sqrt{h}=(\frac{r_0^2}{r_A^2}-\frac{2m}{r_0})^{1/2}r_0^2e^{2f}\sin\theta~.
\end{equation}
Here, the counterterm $I_{ct}$ necessary for canceling the divergence of $I$ is given by the boundary integral
\begin{equation}
I_{ct}=\frac{1}{4\pi r_A}\int d^3x\sqrt{h}=\frac{\beta r_0^2}{r_A}(\frac{r_0^2}{r_A^2}-\frac{2m}{r_0})^{1/2}~.
\end{equation}
Then, in the limit $r_0\rightarrow\infty$, we obtain
\begin{equation}
I=-\frac{\beta}{2}m=-\frac{\pi}{3}(2r_A^2m)^{2/3}~.
\label{action2}
\end{equation}
By virtue of the condition (\ref{condition}) it is clear that no correction due to the distortion function $f$ appears in this formula for $I$. Hence, for the thermodynamic mass energy $E$ and the entropy $S$ defined by \begin{equation}
E=\frac{\partial I}{\partial\beta}~, ~~S=\beta M-I~,
\end{equation}
we arrive at the well-known results
\begin{equation}
E=m~, ~~S=\pi r_H^2=\pi(2r_A^2m)^{2/3}~.
\end{equation}
Thermodynamic properties of distorted AdS black holes become almost the same as SADS black holes in the small $\delta$ domain.

\section{CONNECTION TO BLACK STRINGS}

In the previous section we have seen that the thermodynamic evolution of  black holes to a distorted configuration can occur without changing the mass energy and the entropy. Here we present an explicit example of a family of solutions describing a distortion from a SAdS black hole to a uniform black string which has the cylindrical metric \cite{Lemos1}
\begin{equation}
ds^2=-(\frac{r^2}{r_A^2}-\frac{2m}{r})dt^2+(\frac{r^2}{r_A^2}-\frac{2m}{r})^{-1}dr^2+
r^2(\frac{dz^2}{z_0^2}+d\phi^2) ~ ,
\label{cylinder}
\end{equation}
where $r$ is interpreted as a cylindrical radius, and $z_0$ is an arbitrary parameter with the dimension of length. The parameter $m$ differs from the black hole mass which will become divergent for this cylindrical system with the infinite horizon area. Hence, to obtain such a black string solution, we must consider the increase of the SAdS mass in addition to the cylindrical distortion.

Note that the SAdS metric rewritten into the form
\begin{equation}
ds^2=-(a^2-\frac{2m}{r}+\frac{r^2}{r_A^2})dt^2+(a^2-\frac{2m}{r}+\frac{r^2}{r_A^2})^{-1}dr^2+
\frac{r^2}{a^2}(d\theta^2+\sin^2\theta d\phi^2) ~ ,
\label{sads3}
\end{equation}
represents a spherical black hole with mass $m/a^3$. If the distortion procedure explained in Sec. II is applied to this spherical metric (\ref{sads3}), the distorted metric is given by
\begin{equation}
e^{2\nu}\simeq e^{-2\mu}\simeq \frac{1}{\delta}(x^2-\frac{1}{x})\{1+2\delta\nu_1+a^2\delta(x^2-\frac{1}{x})^{-1}\}~,
\label{h2}
\end{equation}
and
\begin{equation}
e^{2\psi}\simeq \frac{e^{2f(\theta)}}{a^2}~,
\label{f2}
\end{equation}
instead of Eqs. (\ref{h}) and (\ref{f}).
Then it is easily found that
\begin{equation}
\frac{6h}{a^2}-1=e^{-2f}(\frac{d^2f}{d\theta^2}+\cot\theta\frac{df}{d\theta}-1)
\label{basic2}
\end{equation}
as a modified version of Eq. (\ref{basic}). Though the condition (\ref{condition}) is assumed to be preserved, the area of $S^2$ (and the mass $m/a^3$) can increase as the positive parameter $a$ decreases. 

The interesting example of the distortion function $f$ is 
\begin{equation}
e^{2f}=\frac{b}{\sin^2\theta+\gamma}~,
\label{example}
\end{equation}
where $\gamma$ is an arbitrary positive parameter, and the condition (\ref{condition}) leads to
\begin{equation}
b=2\sqrt{1+\gamma}\times(\ln\frac{\sqrt{1+\gamma}+1}{\sqrt{1+\gamma}-1})^{-1}~.
\end{equation}
Further, from Eq. (\ref{basic2}) we obtain
\begin{equation}
\frac{6h}{a^2}=1+\frac{\gamma}{b}[1-\frac{2(1+\gamma)}{\sin^2\theta+\gamma}]~.
\label{h3}
\end{equation}
We find that in the large  $\gamma$ domain (where $a$ is assumed to be $a\simeq 1$) the function $f$ represents a small quadrupole distortion given by 
\begin{equation}
f\simeq\frac{1}{3\gamma}P_2(\cos\theta)~,
\end{equation}
using the approximated relation $b\simeq\gamma+(2/3)$. On the other hand, in the small $\gamma$ domain, we obtain
\begin{equation}
b\simeq(\ln\frac{2}{\sqrt{\gamma}})^{-1}\ll 1~.
\end{equation}
We must keep the metric function $e^{2\psi}$ finite even in the limit $\gamma\rightarrow 0$. Hence, from Eq. (\ref{f2})  the parameter $a$ is chosen to be $a^2\simeq b\rightarrow 0$ in the small $\gamma$ limit, for which it is easy to see that the metric functions given by Eqs. (\ref{h2}) and (\ref{f2}) reduce to the cylindrical metric given by Eq. (\ref{cylinder}), using the coordinate transformation
\begin{equation}
\frac{z}{z_0}=\int_\theta^{\pi/2} \frac{d\theta}{\sin\theta}~.
\end{equation}
In fact, the limit $a^2\rightarrow0$ for the metric (\ref{sads3}) means a transition to the horizon with zero scalar curvature $k=0$. Further, according to Eq. (\ref{h3}) the function $h$ giving the perturbed metric $\nu_1$ also vanishes in the same limit. By virtue of the disappearance of $h$ we obtain a non-perturbed uniform black string solution. Thus, we can claim that the distorted black hole metric written by Eq. (\ref{example}) is a one-parameter family of solutions giving a SAdS black hole in the limit $\gamma\rightarrow\infty$ (i.e., $a\rightarrow 1$) and a uniform black string in the limit $\gamma\rightarrow 0$ (i.e., $a\rightarrow 0$). 

Finally, let us discuss the validity of the Penrose inequality \cite{Bray} and the hoop conjecture \cite{Thorne} for black hole geometry in the distortion process to the black string. The Penrose inequality states that the black hole mass given by $M=m/a^3$ is not less than $\sqrt{A/16\pi}$, where $A$ is the horizon area given by $A=4\pi(2mr_A^2/a^3)^{2/3}$. The parameter $a$ is required to decrease from unity to zero as $\gamma$ decreases to zero. Then the validity of the Penrose inequality is apparent, because the ratio given by
\begin{equation}
\frac{M}{\sqrt{A/16\pi}}=\frac{1}{\delta a^2}
\end{equation}
cannot be smaller than unity in the range $a\leq1$.

On the other hand the hoop conjecture states that any circumference $C$ on the horizon is bounded by $C\leq4\pi M$. Here, the circumference $C$ should be given as the length of a closed loop at a constant $\phi$, namely, we obtain
\begin{equation}
C=\frac{(2mr_A^2)^{1/3}}{a}\int_0^{\pi}e^{f}d\theta~.
\end{equation}
Then, in the limit $\gamma\rightarrow 0$, we can roughly evaluate the ratio $C/4\pi M$ as
\begin{equation}
\frac{C}{4\pi M}\sim \delta(\ln\frac{1}{\gamma})^{-1/2}\ll 1~,
\end{equation}
using $a^2\simeq b\sim 1/\ln(1/\gamma)$. We find that the hoop conjecture remains valid even for the extremely prolonged distortion $\gamma\rightarrow 0$. 

It should be noted that the hoop conjecture becomes consistent with the existence of such a prolonged horizon owing to the rapid increase of mass $M=m/a^3$ as $a^2$ decreases in proportion to $b$. Hence, one may expect the breakdown of the hoop conjecture for a prolonged distortion (i.e., $\gamma\ll1$) keeping $M=m$ (i.e., $a=1$), which gives the ratio
\begin{equation}
\frac{C}{4\pi M}\sim \delta(\ln\frac{1}{\gamma})^{1/2}~.
\end{equation}
However, the circumference $C$ can become larger than $4\pi M$ only for $\ln(1/\gamma)$ larger than $1/\delta^2$. This is the case such that the perturbation $\delta\nu_1\sim\delta h $ which is assumed to be very small in our analysis has the amplitude of the order of  $\delta b\geq 1/\delta\gg 1$. The non-perturbative change of the metric functions $\nu$ and $\mu$ may make the horizon structure disappear, consistently with the hoop conjecture. Unfortunately, it is impossible to find solutions describing the disappearance of the event horizon within the framework of our analysis, in which the distortion of the metric functions $\nu$ and $\mu$ is treated perturbatively, though the spherical surface $S^2$ is non-perturbatively distorted by the function $\psi=f(\theta)$. To construct an extremal state of distorted "static" black holes is an interesting problem to be studied in future works.

In summary, we have succeeded in providing a simple non-perturbative scheme to obtain strongly distorted AdS black holes and showing explicitly the existence of a family of solutions connecting spherically symmetric black holes to uniform black strings. From the leading-order calculations in the large mass domain (i.e., $m\gg r_A$), we have found that the black hole mass is identical with the SAdS mass parameter $m$, and the usual thermodynamic relation between the mass and the entropy holds, irrespective of the large distortion of the metric on $S^2$.

We would like to emphasize that the above-mentioned results have been obtained under the large mass approximation $\delta\equiv r_H/2m\ll1$. If the parameter $\delta$ is not so small, the amplitude of the first-order metric perturbation $\psi_1$ distorting geometry on $S^2$ has the same order as $\nu_1$ and $\mu_1$, as was discussed in \cite{Yos}. Namely, no large distortion of $S^2$ is allowed, unless the metric functions $\nu_0$ and $\mu_0$ of SAdS black holes are also strongly disturbed. In particular, in the small mass domain corresponding to $m\ll r_A$ (i.e., $\delta\simeq 1$), the second-order perturbations $\nu_2$ and $\mu_2$ can become significantly large even if the first-order perturbations $\nu_1$ and $\mu_1$ has the small amplitude of the order of $(m/r_A)^{1/2}$. In relation to the uniqueness theorem (as well as the hoop conjecture) for black hole solutions in four-dimensional $\Lambda=0$ spacetimes, it is a remaining important task to check non-perturbatively whether the horizon distorted from spherical symmetry can exist for the small mass domain.

\begin{acknowledgments}
The author would like to thank Dr. H. Yoshino for helpful discussions.
\end{acknowledgments}

\bibliography{scalar}

\end{document}